\documentclass[useAMS,usenatbib]{mn2e}

\usepackage{graphicx} 

\title[Accretion disc-outflow connection in 3C~111]{X-ray evidence for the accretion disc-outflow connection in 3C~111}
\author[F. Tombesi et al.]{F. Tombesi$^{1,2}$\thanks{E-mail: ftombesi@astro.umd.edu}, R. M. Sambruna$^{3}$, J. N. Reeves$^{4}$, C. S. Reynolds$^{2}$ and V. Braito$^{5}$\\
$^{1}$X-ray Astrophysics Laboratory and CRESST, NASA/Goddard Space Flight Center, Greenbelt, MD 20771, USA\\
$^{2}$Department of Astronomy, University of Maryland, College Park, MD 20742, USA\\
$^{3}$Department of Physics and Astronomy, MS 3F3, 4400 University Drive, George Mason University, Fairfax, VA 22030\\
$^{4}$Astrophysics Group, School of Physical and Geographical Sciences, Keele University, Keele, Staffordshire ST5 5BG, UK\\
$^{5}$Department of Physics and Astronomy, University of Leicester, University Road, Leicester LE1 7RH, UK}
\begin{document}

\date{Accepted ???. Received ???; in original form ???}


\maketitle

\label{firstpage}

\begin{abstract}
We present the spectral analysis of three \emph{Suzaku} XIS observations of 3C~111 requested to monitor the predicted variability of its ultra-fast outflow on $\sim$7~days time-scales. We detect an ionized iron emission line in the first observation and a blue-shifted absorption line in the second, when the flux is $\sim$30\% higher. The location of the material is constrained at $<$0.006~pc from the variability. Detailed modelling support an identification with ionized reflection off the accretion disc at $\sim$20--100$r_g$ from the black hole and a highly ionized and massive ultra-fast outflow with velocity $\sim$0.1c, respectively. The outflow is most probably accelerated by radiation pressure, but additional magnetic thrust can not be excluded. The measured high outflow rate and mechanical energy support the claims that disc outflows may have a significant feedback role. This work provides the first direct evidence for an accretion disc-outflow connection in a radio-loud AGN, possibly linked also to the jet activity.
\end{abstract}


\section{Introduction}

There are several indirect pieces of observational evidence that outflows/jets are coupled to accretion 
discs in black hole accreting systems, from Galactic to extragalactic sizes. Recently, the most direct evidence came from the Galactic microquasar GRS~1915$+$105 (Neilsen \& Lee 2009), where the 
appearance/disappearance of a relativistic disc line and blue-shifted Fe K absorption lines are connected to the state of the source and the radio jet.
In radio-loud AGNs, relativistic jets are routinely observed at radio, optical and X-rays. 
But the first discovery of disc outflows in this class of sources is due to \emph{Suzaku}. A systematic analysis of the 4--10~keV 
spectra of five bright Broad-Line Radio Galaxies (BLRGs) showed significant blue-shifted Fe XXV/XXVI K-shell absorption lines at E$>$7~keV in three sources (Tombesi et al.~2010a), namely 3C~111, 3C~120, 3C~390.3. They imply an origin from highly ionized and massive gas outflowing with mildly relativistic velocities of $v_{out}$$\sim$0.1c.

These characteristics are very similar to those of the Ultra-Fast Outflows (UFOs) previously
observed in Seyferts and quasars (e.g., Chartas et al.~2002, 2003; Pounds
et al.~2003; Braito et al.~2007; Cappi et al.~2009; 
Reeves et al.~2009). In particular, a uniform and systematic search for blue-shifted Fe K absorption 
lines in a large sample of Seyferts observed with XMM-Newton was performed by Tombesi et
al.~(2010b). This allowed the authors to assess their global significance and high detection fraction of $\ga$40\%. 
Tombesi et al.~(2011) then performed a photo-ionization modelling of these absorbers and derived the distribution of the main physical parameters. The outflow velocity is in the range $\sim$0.03--0.3c, with a peak and mean value at $\sim$0.14c, the ionization is in the range log$\xi$$\sim$3--6~erg~s$^{-1}$~cm, with a mean value of $\sim$4.2~erg~s$^{-1}$~cm, and the column density is in the interval $N_H$$\sim$$10^{22}$--$10^{24}$~cm$^{-2}$, with a mean value of $\sim$$10^{23}$~cm$^{-2}$. 
The mass outflow rate of these UFOs can be comparable to the accretion rate and their kinetic power can correspond to fractions of the bolometric luminosity and is comparable to the typical jet power for the radio-loud sources. Therefore, they may play a significant role in the AGN cosmological feedback (e.g., King 2010).

Theoretically, the complex coupling between radiation, magnetic fields and matter that should be considered to properly explain the formation of outflows/winds from accretion discs has not been accurately solved yet. However, simulations show that disc outflows are ubiquitously produced and can be accelerated to velocities $\ga$0.1c by radiation and/or magnetic forces (King \& Fabian 2003; Proga \& Kallman 2004; Oshuga et al.~2009; Fukumura et al.~2010; Sim et al.~2010). The detection of UFOs in BLRGs represents an important step for models of jet formation and for our understanding of the jet-disc connection. From a theoretical perspective, disc outflows are a necessary, although not sufficient, ingredient for jet formation. For instance, in the magnetic tower jet simulations there is the possibility that some plasma is trapped and dragged upward by magnetic field lines and winds/outflows may provide the external pressure needed to effectively collimate the jet (e.g., Kato et al.~2004; McKinney~2006). Thus, models attempting to explain the link between the jet and the accretion process will have to take these components into account. 

The bright BLRG 3C~111 was proposed by us in the \emph{Suzaku} GO5 as the best candidate for a follow-up study of the predicted $\sim$7~days time-scale variability of its UFO, detected by Tombesi et al.~(2010a) in a previous 2008 observation. 3C~111 is a powerful and X-ray bright (2--10~keV luminosity $\sim$$1-4\times 10^{44}$~erg~s$^{-1}$) FRII radio galaxy at $z$=0.0485, with a radio-loudness log$R=$2.35 (Sikora et al.~2007). Its AGN is powered by accretion onto a super-massive black hole (SMBH) with mass\footnote{Chatterjee et al.~(2011) used H$\alpha$ measurements. Marchesini et al.~(2004) assumed the bulge luminosity relation. The discrepancy is probably manily due to the different extinction adopted. We suppose the former is probably more reliable.} $\sim$$(2-30)\times 10^{8}$~$M_{\odot}$ (Chatterjee et al.~2011 and Marchesini et al.~2004, hereafter C11 and M04, respectively). It is important to note that 3C~111 has been recently the subject of an extensive monitoring campaign to study its accretion disc-jet connection (Chatterjee et al.~2011). Indeed, major X-ray dips in the light curve are followed by ejections of bright superluminal knots in the radio jet. A similar campaign showed the same behaviour for 3C~120 (Chatterjee et al.~2009). Both these sources show the presence of UFOs in the X-rays (Tombesi et al.~2010a; Ballo et al.~2011).

\begin{table*}
\centering
\begin{minipage}{150mm}
\caption{Observation log and best-fitting baseline model parameters.}
\begin{tabular}{lccccccccc}
\hline
Obs & Date & Exp & Counts & Flux & $\Gamma$ & E & I & EW & $\chi^2/\nu$\\
\hline
1 & 2010/9/2 & 59 & 72/0.7 & $2.97\pm0.02$ & $1.65\pm0.02$ & $6.38\pm0.04$ & $2.3\pm0.3$ & $40\pm7$ & 1418.5/1380\\
2 & 2010/9/9 & 59 & 93/0.8 & $3.78\pm0.02$ & $1.68\pm0.02$ & $6.39\pm0.02$ & $2.2\pm0.3$ & $33\pm6$ & 1490.4/1469\\
3 & 2010/9/14 & 52 & 79/0.7 & $3.68\pm0.02$ & $1.70\pm0.02$ & $6.40\pm0.02$ & $2.7\pm0.4$ & $38\pm10$ & 1427.0/1410\\
\hline
\end{tabular}
{\em Note.} Columns: observation number; observation date; net XIS exposure in ks; 4--10~keV XIS~03 source/background counts in units of $10^3$; 4--10~keV flux in units of $10^{-11}$~erg~s$^{-1}$~cm$^{-2}$; photon index; Fe K$\alpha$ rest-frame energy in units of keV; intensity in units of $10^{-5}$~ph~s$^{-1}$~cm$^{-2}$; equivalent width in units of eV; best-fit statistic.
\end{minipage}
\end{table*}

  \begin{figure}
  \centering
   \includegraphics[width=4.6cm,height=7.7cm,angle=-90]{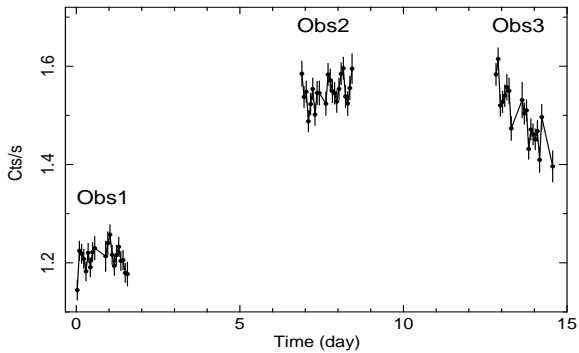}
   \caption{Combined XIS~03 light curves of the three observations in the 4--10~keV band, binned to the satellite orbital period of 5760s.}
    \end{figure}

\section[]{Data analysis and results}

The details of the three \emph{Suzaku} observations of 3C~111 are summarized in Table~1. They are spaced by $\sim$7~days and we refer to them as Obs1, Obs2 and Obs3, respectively. The data were processed using version 2 of the \emph{Suzaku} pipeline and the light curves and spectra were extracted following the standard procedure\footnote{http://heasarc.gsfc.nasa.gov/docs/suzaku/analysis/abc/}. Source spectra were extracted from circles of 3{\arcmin} radius centered on 3C~111, while the background ones were selected from circles of the same size offset from the source. 
In this letter we focus on the spectral analysis of the X-ray Imaging Spectrometer (XIS) in the Fe K band, E$=$3.5--10.5~keV. The analysis was carried out using the \emph{heasoft} v.~6.5.1 package and \verb+XSPEC+ v.12.6.0. The spectra were binned to a minimum of 25 counts to apply the $\chi^2$ minimization in the model fitting. Errors are at the 1$\sigma$ level for one interesting parameter and the line energies are at rest-frame, if not otherwise stated.
The front illuminated XIS~0 and XIS~3 spectra were combined after checking that the continuum and flux agree within 2\% (hereafter XIS~03). The data from the back illuminated XIS~1 chip were checked separately only for consistency. The combined XIS~03 light curves are reported in Fig.~1. We note a $\sim$30\% flux variability between Obs1 and Obs2. 

A power-law continuum with $\Gamma$$\sim$1.65--1.7 and a narrow (unresolved) Fe K$\alpha$ emission line at 6.4~keV provide a good representation of the three spectra (e.g., Reynolds et al.~1998). The narrow emission line is most probably associated to reflection off material located in the outer accretion disc or the broad line region (Chatterjee et al.~2011; Ballo et al.~2011). We include a Galactic absorption fixed to $N_H$$=$$3\times 10^{21}$~cm$^{-2}$ (Kalberla et al.~2005). We refer to these as the baseline models and the best-fitting parameters are reported in Table~1.
From panels 1 and 2 of Fig.~2 we can note the presence of additional emission/absorption lines. This is confirmed in panel 3 of Fig.~2 by the contour plots derived with respect to the baseline models (e.g., Tombesi et al.~2010a). In particular, the contours suggest the presence of an emission line at E$\simeq$6.88~keV in Obs1 and an absorption line at E$\simeq$7.75~keV in Obs2. We initially modelled these features with Gaussian lines. Their best-fitting parameters are reported in Table~2. The fit improvement with respect to the baseline models are $\Delta\chi^2$$=$25.6/15.3 for 3/2 additional model parameters, for the emission/absorption line, respectively. Their detection significance is high, $>$99.99\% and 99.95\% from the F-test and $>$99.9\% and 99.8\% from 1000 Monte Carlo simulations, respectively. The Monte Carlo simulations were performed following the method of Tombesi et al.~(2010a). We estimated the random detection probability for lines in the intervals  E$=$6.5--7.5~keV and E$=$7--10~keV for the emission and absorption feature, respectively. 

We checked that the emission and absorption lines are independently detected in both XIS~0 and XIS~3 and the results are consistent with the XIS~1.
As reported in Table~1, the 4--10~keV background level is $\la$1\% of the source counts. However, we note that the XIS background has an instrumental emission line at E$=$7.5~keV due to Ni K$\alpha$ (Yamaguchi et al.~2006). This is not consistent with but close to the observed energy of the absorption line in Obs2 and we performed several tests to assure that the results are not affected by an erroneous background subtraction. 
We avoided regions too close to the chip corners, where the instrumental contamination is higher, and checked that the results are consistent selecting different background regions. We also inspected that the observed energy of the background line is not consistent at the 90\% with that of the absorption line and the intensity is only $\sim$30\%. We checked that the absorption line parameters are consistent with and without background subtraction and that it is present also fitting the XIS~0 and XIS~3 separately. 
Moreover, given that the background level and instrumental emission line are identical for the three observations and also the source spectra are comparable, if the absorption line was due to some systematics, it would have been present in all the spectra, and especially in Obs1 which has the lowest flux.

Only highly ionized Fe K emission lines are expected in the interval 6.5--7~keV, in particular Fe~XXV He$\alpha$ at 6.7~keV and Fe~XXVI Ly$\alpha$ at 6.97~keV (Kallman et al.~2004). However, the energy of the Gaussian emission line in Obs1 is E$=$6.83--6.93~keV at the 90\% level and therefore is not consistent with these transitions, if not requiring an energy blue/red-shift. The emission line is resolved with $\sigma$$=$$84\pm38$~eV. Therefore, we tested the possible modelling as a blend of rest-frame narrow Fe~XXV--XXVI lines. However, this provides a much worse fit ($\Delta\chi^2/\Delta\nu$$=$17/2) compared to that with a single Gaussian line profile.
An alternative and intriguing possibility is the identification with the blue peak of a relativistic line (e.g., Sambruna et al.~2011). Therefore, we replaced the emission line in Obs1 with the \emph{relline} model of Dauser et al.~(2010). Given the limited S/N, the black hole spin, emissivity and outer radius can not be constrained and were fixed to the typical values of $a$$=$0, $\beta$$=$$-3$ and $r_{out}$$=$1000~$r_g$. We obtain comparable fits assuming in turn an energy of the line equal to the neutral Fe K$\alpha$, Fe~XXV He$\alpha$ or Fe~XXVI Ly$\alpha$. The best fitting parameters in the three cases are $r_{in}$$<$13~$r_g$ and $i$$=$$44^{\circ}\pm2^{\circ}$, $r_{in}$$=$$9\pm2$~$r_g$ and $i$$=$$32^{\circ}\pm2^{\circ}$, $r_{in}$$=$$57^{+33}_{-16}$~$r_g$ and $i$$=$$13^{\circ}\pm4^{\circ}$, respectively.  The fit improvement is $\Delta\chi^2/\Delta\nu$$=$26/2, which corresponds to a confidence level $>$99.99\%. However, if we require the disc inclination to be consistent with that of the jet of $\sim$18$^{\circ}$ (Jorstad et al.~2005) or $10^{\circ}<i<26^{\circ}$ (Lewis et al.~2005), the solutions with ionized lines are preferred. The Fe~XXVI model is shown in Fig.~2 (Panel 4). In particular, if we adopt an inclination of $i$$=$$18^{\circ}$ and let the energy free to vary, we obtain E$=$$6.94\pm0.05$~keV, which is consistent with Fe~XXVI, and $r_{in}$$=$$20^{+10}_{-5}$~$r_g$. The fit improvement is still high, $\Delta\chi^2/\Delta\nu$$=$25/2. The intensity of the line is $I$$=$$(2.4\pm0.6)\times 10^{-5}$~ph~s$^{-1}$~cm$^{-2}$ and EW$=$$54\pm27$~eV. 


The energy of the absorption line at 7.75~keV in Obs2 is also not consistent with any known atomic transition. Assuming a conservative identification with the closest ones due to Fe~XXV--XXVI, the line energy requires a blue-shifted velocity of $\sim$0.1-0.14c. We modelled the absorber with the photo-ionization code {\sc Xstar} (Kallman \& Bautista 2001). We assumed a standard $\Gamma$$=$2 power-law continuum with cut-off at 100~keV and standard Solar abundances (Asplund et al.~2009). Due to the limited energy resolution of the XIS, the absorption line is not resolved, but we can place a 90\% upper limit of 10,000~km/s. Therefore, we performed fits using three {\sc Xstar} tables with turbulent velocities of 1000~km/s, 3000~km/s and 5000~km/s and averaged the resultant fit parameters. To search for the possible best-fitting solution we stepped through the absorber redshift in small increments of $\Delta z$$=$$10^{-3}$ in the interval between $0.1$ to $-0.4$, leaving also the other parameters of the absorber, i.e. $N_H$ and log$\xi$, and the continuum free to vary. A slightly better fit is provided by the 3000~km/s table, which is shown in Fig.~2 (Panel 4). The best-fitting solution yields $v_{out}$$=$$0.106\pm0.006$c, log$\xi$$=$$4.32\pm0.12$~erg~s$^{-1}$~cm and $N_H$$=$$(7.7\pm2.9)\times 10^{22}$~cm$^{-2}$. This is consistent with Fe~XXVI being the dominant ionic species. The fit improvement is high, $\Delta\chi^{2}/\Delta\nu$$=$23/3, which corresponds to a detection probability $\ge$99.99\%.

  \begin{figure*}
  \centering
   \includegraphics[width=13cm,height=6cm,angle=0]{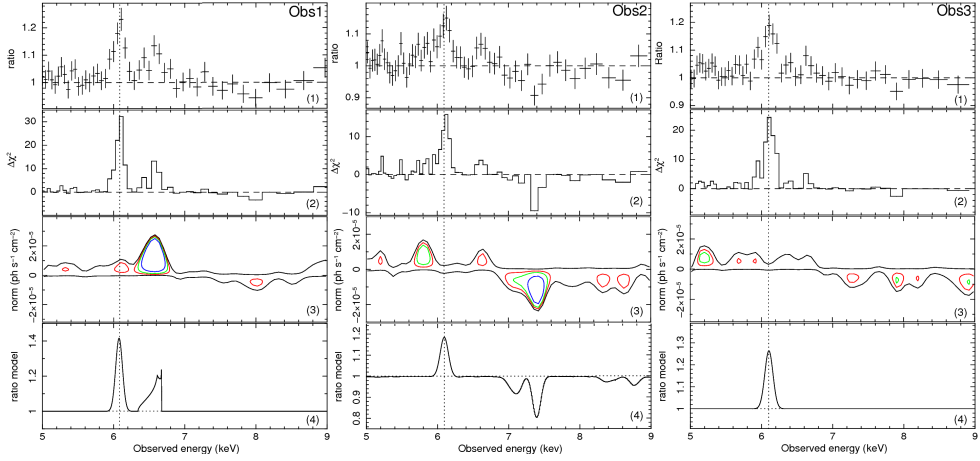}
   \caption{Comparison of the XIS~03 spectra of the three observations zoomed in the 5--9~keV interval. \emph{Panel 1:} ratios with respect to a Galactic absorbed power-law continuum. Data were binned to a S/N$=$30 only for plotting. \emph{Panel 2:} $\Delta \chi^2$ residuals. \emph{Panel 3:} contour plots with respect to the baseline models including also the narrow Fe K$\alpha$ emission line. The contours refer to $\Delta\chi^2$ levels of $-2.3$, $-4.61$ and $-9.21$, which correspond to F-test confidence levels of 68\% (red), 90\% (green) and 99\% (blue), respectively. The black indicates the reference level $\Delta\chi^2=+0.5$. \emph{Panel 4:} final best-fitting models. Besides the baseline models they include a relativistic Fe~XXVI emission line in Obs1 and an {\sc Xstar} photo-ionization grid with turbulent velocity of 3000~km/s in Obs2, respectively. See text for more details. The vertical lines indicate the observed energy of the narrow Fe K$\alpha$ emission line.}
    \end{figure*}

The 90\% upper/lower limits of the intensity and EW of the two additional emission and absorption lines in the three observations are reported in Table~2. A constant behaviour of the lines can be tested by fixing the energy and width to the best-fitting values and comparing with the fits with free intensity/EW. We can rule out a constancy of the emission line in Obs2 and Obs3 at 99.7\%. Instead, the constancy of the absorption line in Obs1 and Obs3 is ruled out at 99.9\% and 92\%, respectively. Therefore, both features are significantly variable. Finally, consistent line parameters are found when the broad-band spectra are considered. The complete broad-band analysis and fits including ionized reflection models, along with a comparison with previous X-ray (e.g., Tomesi et al.~2010a; Ballo et al.~2011) and radio observations of 3C~111 (e.g., Chatterjee et al.~2011) will be presented in forthcoming papers.

\section[]{Discussion and conclusions}


From the spectral variability time-scales of $\sim$7~days we can constrain the distance of the material(s) producing the emission/absorption features from the X-ray source to $d<c \Delta t< 1.8\times 10^{16}$~cm ($<$0.006~pc), which corresponds to $<$40--600~$r_g$ assuming the black hole mass of M04 or C11, respectively. This independently tells us that the material must be rather close to the SMBH.
Phenomenologically, the emission line at 6.88~keV in Obs1 is in a region of the spectrum in which only emission lines from Fe~XXV--XXVI would be expected, which corresponds to log$\xi$$\sim$3--4~erg~s$^{-1}$~cm (Kallman et al.~2004). The ionization parameter is defined as $\xi$$=$$L_{ion}/nr^{2}$ (Tarter et al.~1969), where $L_{ion}$ is the 1--1000~Ryd (1~Ryd$=$13.6~eV) absorption corrected ionizing luminosity, $n$ is the number density and $r$ the distance from the source. Substituting $L_{ion}$$\simeq$$5.8\times 10^{44}$~erg/s for Obs1 extrapolated from the spectrum and the distance $d$, we obtain a density $n$$>$$10^{9}$~cm$^{-3}$. Given the compactness of the distribution and substituting in $N_H$$\sim$$nd$, we can estimate that the reflecting material is possibly Compton thick, with $N_H$$\ga$$10^{25}$~cm$^{-2}$.
However, the measured line energy is not consistent with Fe~XXV--XXVI at rest. 
An association with Fe~XXV implies a blue-shift of $\sim$0.026c. This could be explained by the bulk of reflection from the approaching side of the disc or a large scale, Compton-thick outflow. However, the latter seems unlikely given that we do not observe a P-Cygni profile and simultaneous absorption lines.
Instead, an identification with Fe~XXVI implies a red-shift of $\sim$0.013. If interpreted as a gravitational redshift, this indicates that the bulk of the emission is at $\la$80~$r_g$. 
Therefore, both phenomenological arguments and a detailed modelling with a relativistic profile suggest the line in Obs1 is most probably generated by reflection off the accretion disc at $\sim$20--100$r_g$ from the SMBH.

Concerning the blue-shifted absorption line in Obs2, the photo-ionization modelling indicates that the absorber is highly ionized, log$\xi$$\simeq$4.3~erg~s$^{-1}$~cm, possibly mildly Compton-thick, $N_H$$\simeq$$8\times 10^{22}$~cm$^{-2}$, and has a mildly relativistic outflow velocity $v_{out}$$\simeq$0.1c. This clearly suggests an association with a UFO (Tombesi et al.~2010a,b). Following similar reasonings to those in Tombesi et al.~(2010a), we can estimate several physical parameters of the outflow. Throughout, we assume a covering fraction $C$$\sim$0.5 as estimated for the UFOs in Tombesi et al.~(2010a, b). The absorption corrected ionizing luminosity in Obs2 extrapolated from the spectrum is $L_{ion}$$\sim$$8.1\times 10^{44}$~erg/s. The $\sim$7~days variability indicates a compact absorber. Substituting the best-fit parameters in the $\xi$ definition, we obtain a distance from the ionizing source $<$$5\times 10^{17}$~cm, consistent with the variability upper limit $d$. On the other hand, substituting the distance $d$, we can estimate a density $n$$\ga$$10^{8}$~cm$^{-3}$. 

The bolometric luminosity can be derived from the relation $L_{bol}$$\simeq$$10L_{ion}$ (McKernan et al.~2007). Assuming a constant flow, the mass outflow rate is $\dot{M}_{out}$$=$$4\pi C \frac{L_{ion}}{\xi} m_H v_{out}$$\sim$1~$M_{\odot}/yr$ and the mass accretion rate is $\dot{M}_{acc}$$=$$L_{bol}/\eta c^2$$\sim$1~$M_{\odot}/yr$, using $\eta$$=$0.1. Then, $\dot{M}_{out}/\dot{M}_{acc}$ $\sim$1, suggesting that the instantaneous ejection of mass in Obs2 is comparable to that accreted by the SMBH in that moment. The $\sim$40\% increase in luminosity between Obs1 and Obs2 could have been driven by an increase in efficiency and/or accretion rate. Interestingly, from the ratio $\frac{\dot{M}_{out}v_{out}}{L_{bol}/c}$$\sim$1 we derive that the wind momentum is of the order of the radiation field or photon momentum and therefore radiation pressure could have played an important role in accelerating the wind (e.g., King \& Pounds 2003; King 2010). 
The mechanical power of this outflow is $\dot{E}_{K}$$=$$\frac{1}{2} \dot{M}_{out} v_{out}^{2}$$\sim$$5\times 10^{44}$~erg/s, which is comparable to the typical jet power for radio galaxies of $\sim$$10^{43}$--$10^{45}$~erg/s (Rawlings \& Saunders 1991). Then $\dot{E}_{K}$$\sim$0.06$L_{bol}$, again consistent with momentum driven outflows (e.g., King \& Pounds 2003; King 2010). 
The Eddington luminosity and relative accretion rate are $L_{Edd}$$\sim$$2.6$--$39.0\times 10^{46}$~erg/s and $\dot{M}_{Edd}$$=$$L_{Edd}/\eta c^2$$\sim$5--50~$M_{\odot}/yr$, depending on the SMBH mass estimates of C11 or M04. The Eddington ratio $L_{bol}/L_{Edd}$$\sim$$\dot{M}_{acc}/\dot{M}_{Edd}$$\la$0.3 is moderate. However, the $L_{Edd}$ should be regarded only as an upper limit for the luminosity needed to accelerate winds by radiation pressure because it takes into account only the opacity due to Thomson scattering. Considering only this term, we obtain a lower limit to the wind opacity in Obs2 of $\tau$$\sim$$\sigma_{T} N_H$$\sim$0.05. 

It is important to note that in Obs2 we observe the wind when it has already been accelerated at $\sim$0.1c and, due to the low opacity and high ionization, it follows that it would be hard to further accelerate it through radiation pressure alone. However, the ratio of $\sim$1 between the wind and photon momenta suggests that radiation pressure had an important role in launching the outflow. 
Given the high column density of $\sim$$10^{23}$~cm$^{-2}$ in Obs2, it is plausible that during the launching and acceleration process the density was higher and therefore the wind was less ionized and possibly optically thick. In this case, the many absorption lines and edges from light elements up to iron could have provided the additional opacity (e.g., Proga \& Kallman 2004). 
Imposing $\tau$$\sim$1 we can estimate the photospheric radius (King \& Pounds 2003; Pounds et al.~2003), which indicates the region where the photon momentum was deposited into the wind, and we derive $R_{ph}$$\sim$100~$r_g$. 
We note that the observed velocity of $\sim$0.1c and column density are only conservative estimates, corresponding to the components along our line of sight. The first value corresponds to the escape velocity at $\sim$100--200~$r_g$ and, thus, the outflow is likely to leave the system even if does not undergo further acceleration. Moreover, the outflow velocity and $\sim$7~days variability are comparable to the disc Keplerian velocity and dynamical time scale at that location.

Therefore, we can derive a picture in which an increase in the X-ray illumination, possibly linked to a rise in efficiency/accretion rate, causes a truncation of the inner parts of the disc or overionizes the material responsible for the relativistic line. This can explain the lack of the 6.88~keV emission line in Obs2--3. Then, an outflow is lifted up from the disc surface at $\sim$100$r_g$ and is accelerated by radiation pressure to a velocity of $\sim$0.1c between Obs1--2. The outflowing plasma might then flow along the magnetic field lines and be further accelerated by magnetic torque and pressure (Kato et al.~2004; Oshuga et al.~2009; Fukumura et al.~2010). This latter consideration is plausible given the low inclination of the outflow with respect to the jet of $\sim 18^{\circ}$ and the fact that 3C~111 is a well known superluminal source. In particular, this ionized outflow might be associated with external layers of the jet and provide the required external pressure to collimate it (Kato et al.~2004; McKinney~2006). The possible disappearence of the blue-shifted absorption line in Obs3 could then be explained by an increase in ionization and/or the absorber moving away from the line of sight. 
The sequence X-ray dip followed by ejection event is reminiscent of the disc-jet connection recently found in this source by Chatterjee et al.~(2011). However, the actual link between the outflow and jet is not clear yet. If the corona, where the X-rays originate, is indeed related to the base of the jet, then the loading of electrons in the jet may cause a flux decrease. Then the particles are released through an outflow and an increase in flux follows. In a subsequent paper we will report on a more detalied comparison with respect to the radio properties of this source and possibly the connection between the accretion disc, outflows and the jet would be better quantified.

\begin{table}
\caption{Best-fitting parameters of the additional emission and absorption lines.}
\begin{tabular}{lcccc}
\hline
Obs & E & $\sigma$ & $I$ & EW \\
\hline
1 & $6.88(6.56)\pm0.03$ & $84\pm38$ & $+1.7\pm0.3$ & $+35\pm8$ \\
  & $\equiv$$7.75(7.39)$ & $\equiv$$10$ & $>-0.4$ & $>-13$ \\
2 & $\equiv$$6.88(6.56)$ & $\equiv$$84$ & $<+0.9$ & $<+16$ \\
  & $7.75(7.39)\pm0.03$ & $\equiv$$10$ & $-1.2\pm0.3$ & $-26\pm6$ \\
3 & $\equiv$$6.88(6.56)$ & $\equiv$$84$ & $<+0.9$ & $<+16$ \\
  & $\equiv$$7.75(7.39)$ & $\equiv$$10$ & $>-1.0$ & $>-19$ \\ 
\hline
\end{tabular}
{\em Note.} Columns: observation; rest-frame (observed) energy in keV; line width in eV; intensity in units of $10^{-5}$ ph~s$^{-1}$~cm$^{-2}$; EW in eV. Lower/upper limits are at the 90\% level.
\end{table}

\section*{Acknowledgments}

FT thank A. P. Marscher for the useful discussion. CSR would like to thank NASA for support under  Suzaku Guest Observer grant NNX10AR31G.

\end{document}